# EVENT CATEGORIES IN THE EDELWEISS WIMP SEARCH EXPERIMENT

A. Benoit[a], L. Bergé[b], A. Broniatowski[b], B. Chambon[c], M. Chapellier[d], G. Chardin[e], P. Charvin[e,f], M. De Jésus[c], P. Di Stefano[e*], D. Drain[c], L. Dumoulin[b], P. Garoche[g], J. Gascon[c], C. Goldbach[h], M. Gros[e], A. Juillard[b], A. de Lesquen[e], D. L'Hôte[d], J. Mallet[e], J. Mangin[i], S. Marnieros[b], N. Mirabolfathi[b], L. Miramonti[e], L. Mosca[e], X-F. Navick[e], G. Nollez[h], P. Pari[d], S. Pécourt[c], E. Simon[c], M. Stern[c], J-P. Torre[j]

[a]*Centre de Recherche sur les Très Basses Températures, BP 166, 38042 Grenoble, France*
[b]*CSNSM, IN2P3-CNRS, Univ. Paris XI, bat. 108, F-91405 Orsay Cedex, France*
[c]*IPN Lyon and UCBL, IN2P3-CNRS, 43 Bd. du 11 novembre 1918, F-69622 Villeurbanne Cedex, France*
[d]*CEA, Centre d'Etudes Nucléaires de Saclay, DSM/DRECAM, F-91191 Gif-sur-Yvette Cedex, France*
[e]*CEA, Centre d'Etudes Nucléaires de Saclay, DSM/DAPNIA, F-91191 Gif-sur-Yvette, France*
[f]*Laboratoire Souterrain de Modane, CEA-CNRS, 90 rue Polset, F-73500 Modane, France*
[g]*Laboratoire de Physique des Solides, Université Paris-Sud Orsay, 91405 Orsay, France*
[h]*Institut d'Astrophysique de Paris, INSU-CNRS, 98 bis Bd. Arago, F-75014 Paris, France*
[i]*LPUB, Université de Bourgogne, 21078 Dijon, France*
[j]*Service d'Aéronomie, BP 3, 91371 Verrières le Buisson Cedex, France*

**Abstract**

Four categories of events have been identified in the EDELWEISS-I dark matter experiment using germanium cryogenic detectors measuring simultaneously charge and heat signals. These categories of events are interpreted as electron and nuclear interactions occurring in the volume of the detector, and electron and nuclear interactions occurring close to the surface of the detectors (within ≈10-20 µm of the surface). We discuss the hypothesis that low energy surface nuclear recoils, which seem to have been unnoticed by previous WIMP searches, may provide an interpretation of the anomalous events recorded by the UKDMC and Saclay NaI experiments. The present analysis points to the necessity of taking into account surface nuclear and electron recoil interactions for a reliable estimate of background rejection factors.

## Introduction

The quest for non baryonic Dark Matter under the form of Weakly Interacting Massive Particles (WIMPs) faces the challenge of extremely low interaction rates (< 1 evt/kg/day) for realistic SUSY models. For kinematical reasons, WIMP interactions at detectable energies (> 1 keV) are expected to occur primarily with nuclei of the target material, by opposition to the radioactive background, which involves mainly electron recoils. Detectors with extremely low radioactive background [1] and/or detectors with highly efficient discrimination against electron recoils are then required for this search. Various experimental methods have been proposed, and in some occasions implemented in large-scale experiments, to distinguish between nuclear and electron recoils [2-8]. In particular, NaI-based experiments [3-5] use the property of faster scintillation decay times for nuclear recoils compared to electron recoils to discriminate between electron and nuclear recoils. On the other hand, cryogenic experiments have used the simultaneous measurement of heat and charge [6-10], or of light and heat signals [11, 12] to efficiently reject the gamma and electron radioactive background.

In the following, the data recorded by the EDELWEISS experiment in its prototype germanium detectors are used to evidence four categories of events, instead of the two initially expected populations of electron and nuclear recoils.

---

* Present address : Max-Planck-Institut für Physik, Föhringer Ring 6, D-80805 München, Germany

## The EDELWEISS experiment

EDELWEISS is a Dark Matter WIMP direct detection experiment set in the Fréjus underground laboratory, adjacent to a highway tunnel connecting France and Italy. The experimental setup of the EDELWEISS-I experiment has been described elsewhere [13]. In initial data takings, data have been accumulated using a 70 g (48 mm diameter, 8 mm thick) high purity Ge ionization-heat detector [14] without its Roman lead shielding and without radon removal. Most of these data have been recorded under charge collection voltages of 2 volts or 6 volts [10, 14]. The heat and ionization channels exhibit energy resolutions of $\approx 1$ and $\approx 1.2$ keV FWHM at 122 keV ($^{57}$Co γ-source), respectively. In the following, these data have been used to characterize the different event categories recorded by the ionization-heat detectors.

## Nuclear recoil and electron recoil calibrations

The experimental response of our germanium detectors to electron and nuclear recoils has been measured using radioactive sources. Electron recoils have been calibrated using radioactive $^{60}$Co and $^{57}$Co sources. The $^{60}$Co source, with its two 1.173 and 1.333 MeV gamma-ray lines, provides a rather uniform exposure of the 70 g germanium detectors. The scatter diagram of the charge and heat signal amplitudes recorded during one of these calibration runs under a charge collection bias of 6 Volts is shown in Fig. 1 and provides a reference for volume electron interactions. Calibrations using a $^{57}$Co source (gamma-ray energies of 122 and 136 keV) have also been realized, and fluorescence X-ray lines produced, e.g., by the archeological lead present in some calibration runs and close to the detector. These runs have been used to test the linearity of the detector response. The scatter diagram corresponding to the charge-phonon calibration of a 70 g germanium detector under a charge collection bias of 6 Volts using such a $^{57}$Co γ-ray source is shown in Fig. 2. Using these data, the charge-phonon response of the detector in this energy range has been found to be linear to a precision better than 1 keV over the [0, 136 keV] energy interval. In Fig. 2, it can be seen that a few percent of the interactions lead to off-axis charge-phonon ratios.

The response of our detectors to nuclear recoils has been measured using a $^{252}$Cf neutron source. The associated gamma radiation emitted by the source has been strongly reduced by the use of a lead shield and most interactions are due to neutron induced nuclear recoil interactions. Due to the interaction length of neutrons in germanium in the MeV range (of the order of 10 cm at these energies), a significant fraction of the interactions (typically 25%) are expected to result from multiple scattering interactions of the neutron inside the detector. The scatter diagram of the charge and heat signal amplitudes recorded during the neutron calibration runs, under the same charge collection bias of 6 Volts, is shown in Fig. 3. It can be seen that a clear separation between electron and nuclear recoil events can be realized on an event by event basis down to an energy of $\approx 4$ keV electron-equivalent (e.e.). A detailed study of the rejection factors achieved with these prototype detectors will be found elsewhere [15].

## Recoil energy determination

In Figs. 1-3, the charge and phonon amplitudes have been plotted for each interaction. However, the phonon amplitude is not directly proportional to the recoil energy since it includes a contribution from the Joule heating resulting from the charge drift towards the electrodes, the so-called Luke-Neganov effect [16]. The recoil energy $E_{rec}$ has to be derived from the charge and phonon amplitudes $E_{ch}$ and $E_{ph}$ using the formula [16, 17]:

$$E_{rec} = E_{ph}\left(1 + \frac{eV}{\varepsilon_\gamma}\right) - E_{ch}\frac{eV}{\varepsilon_\gamma},$$

where $V$ is the bias voltage for charge collection, $e$ is the elementary charge and $\varepsilon_\gamma$ is the average energy required to produce an electron-hole pair (in eV) for an electron recoil. The Luke-Neganov effect can be determined experimentally by considering events of a given energy with incomplete charge collection. Such events can be observed on Fig. 2, where the Pb fluorescence line doublet at an energy of 72.80 and 74.97 keV is seen to present a charge leakage band leading to a charge amplitude dependence for the phonon channel. This phonon amplitude which reads [17]

$$E_{ph} = \frac{eV}{eV + \varepsilon_\gamma} E_{ch} + \frac{\varepsilon_\gamma}{eV + \varepsilon_\gamma} E_0$$

can then be extrapolated to zero charge in order to recover the recoil energy. Under a bias voltage of 6 volts, it is then found experimentally that ≈68 % of the phonon amplitude of an electron recoil is due to the Luke effect induced by the charge drift. This value is consistent with the standard value of $\varepsilon_\gamma \approx 3$ eV for 77K germanium detectors [18].

## Electron surface interactions and volume nuclear recoils

Initial runs, for a total exposure of 1.17 kg × day, of our first 70 g Ge detector [10] have evidenced a population of events intermediate between the electron recoil and the nuclear recoil calibration lines (Fig. 4). This population is attributed to incomplete charge collection of surface events: approximately 50 % of the charge initially produced in an interaction close to the surface is lost since charge carriers diffusing against the electric force towards the nearest electrode are captured by this electrode and cannot contribute to the electrical signal (see also [19]). It can be seen in Fig. 4 that the number of surface electron recoil events is comparable at low energies to that of volume electron recoil events. These events may be associated with a small tritium contamination resulting in low energy beta decay electrons or with soft X-rays emitted by the fluorescence of surrounding materials (copper and germanium, e.g.).

In the region of the charge-phonon scatter diagram where nuclear recoils are expected (the precise definition of this region will be discussed in Ref. 15), approximately 20-30 events are observed. These events are not sufficiently separated from the population of electron surface events to be attributed with certainty to nuclear recoil events. On the other hand, the flux and energy spectrum of fast neutron induced nuclear recoils, originating for the major part from radioactivity in the surrounding rock, have been measured in the Fréjus underground laboratory using conventional scintillation detectors. The fast neutron flux was found to be $4.0 \pm 0.2 \; 10^{-6}$ cm$^{-2}$ s$^{-1}$ [20]. The measured flux and energy spectrum of neutrons have been used in a Monte-Carlo simulation to predict the expected number of neutron-induced nuclear recoils in our 70 g Ge detectors. In the absence of paraffin shielding against neutrons, the expected number of interactions above 20 keV recoil energy is $30 \pm 2$ events (statistical error only) for the 1.17 kg × day data set. Taking into account the limited reliability of neutron simulations, it appears that a large fraction, and possibly all the events observed in the nuclear recoil region can be attributed to neutron interactions inside the detector.

## Surface nuclear recoils

In Fig. 4, an initially unexpected category of events can be observed at recoil energies of the order of 80-100 keV and at low charge yield (≈10% of that expected for an electron recoil of the same energy). This group of events is attributed to surface nuclear interactions.

In germanium detectors, for which the material is usually extremely radiopure, most of the surface nuclear interactions are expected to be due to alpha contaminants adsorbed on the surface of the detector. When the alpha particle is detected, its energy of a few MeV makes it impossible to be confused with a low energy WIMP interaction. On the other hand, when an unstable heavy nucleus decays through an alpha disintegration at the surface of the detector, the alpha particle may leave the detector unscathed and, in this case, the energy detected is due to the recoil of the remaining part of the heavy nucleus. Therefore, in typically 50% of the alpha decays, e.g. in the polonium α disintegration:

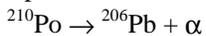

the heavy nucleus fragment $^{206}$Pb recoils with an energy of the order of ≈5 MeV × 4/216 ≈100 keV, by momentum conservation in the alpha disintegration.

In conventional germanium detectors, most of these events are not observed due to the passivation of the detector surface. In a cryogenic detector, however, such a surface passivation over a thickness of several hundreds of microns is hardly possible due to the large additional heat capacity. Surface nuclear interactions are rather easily detected in the phonon channel since their energy lies in the 100 keV range. On the other hand, the charge amplitude of these events is usually much lower: the quenching factor, defined by reference to an electron recoil, of the heavy recoiling nucleus is ²0.25 [21] for volume interactions, and a surface nuclear interaction results, similarly to surface electron interactions, in a ≈50% further loss in the collected charge. These surface nuclear events have also been studied by the IAS group [22] and by the Milano group [23] in the context of measurements of the quenching factor of nuclear recoil events in crystal bolometers without charge measurement.

The electron and nuclear surface events represent a potential threat to the nuclear vs. electron recoil discrimination technique since badly collected or reconstructed surface events may mimic a nuclear recoil event. Therefore, in addition to the passive shielding used in the EDELWEISS experiment, several approaches are being tested to reject or reduce the number of such surface interactions. In particular, preliminary experiments have shown that the interaction point can be partially localized by using the pulse shape of the phonon signal and the relative amplitudes in separate thin film sensors [24]. In addition, a pulse shape discrimination technique using the time structure of the charge signal has also been used to localize the interaction [25].

## Nuclear surface events and anomalous events reported by the UKDMC and Saclay experiments

At present, taken at face value, the experiment reporting the best sensitivity for WIMP detection is the DAMA experiment [3] which claims evidence for an annual modulation in its data under the assumption of a spin-independent interaction. Assuming a spin-dependent WIMP-nucleon interaction, the DAMA experiment reports strong reduction of the background rate using a pulse shape discrimination (PSD) analysis. However, for a WIMP mass in the 100 GeV range, most WIMP interactions are expected to occur at visible energies between a few keV and a few tens of keV. Even with efficient light collection, the small number of photoelectrons detected in present large size NaI crystals makes it impossible to discriminate the events individually below ≈20 keV, where most of the WIMP interactions are expected to show up. Instead, either global light time distributions or averaged pulse shapes are analyzed in terms of two components of electron and nuclear recoils [3-5]. It appears therefore essential that all the existing populations of events be identified and that no unexpected populations exist which could at low energies appear as a linear combination of nuclear and electron recoil distributions.

Over the last two years, however, two of the NaI experiments [4, 5, 26] have gradually realized that (at least) a third unexpected category of events was present in their respective data sets. These events have been named "bump events" by the UKDMC experiment, by reference to the decay time distribution of the scintillation light presenting an enhancement at smaller times, and "U" (unknown) events by the Saclay group.

We suggest here that these anomalous events could be related to the low energy surface nuclear interactions discussed above. To justify this hypothesis, we note that both electron and nuclear surface events have been found experimentally to present scintillation decay times differing from that of volume electron and nuclear recoils. More precisely, the UKDMC experiment has noted that surface alpha interactions of degraded energies exhibit shorter scintillation decay times than nuclear recoils [26]. Similarly, low energy X-ray surface events have been observed by the Saclay group to present decay times shorter than bulk gamma-ray interactions [5]. Furthermore, the rate observed by the UKDMC and Saclay groups, of the order of 1 event/kg/day, appears to be consistent with a fraction of their total alpha contamination (up to the trigger efficiency, the rate of these events should be roughly equal to that of the surface alpha contamination). Finally, the energy detected in a NaI detector for a heavy recoiling nucleus will be in the low energy range where WIMPs might be expected to contribute since the low quenching factor of such heavy recoiling nuclei (measured, e.g. to be ≈ 0.08 for iodine [5]) will result in an energy of a few keV e.e. as observed in a NaI detector.

The energy range of the anomalous events has been measured by the UKDMC experiment to extend to several tens of keV [4, 26], but this can probably be explained in terms of a small fraction of the energy of the (almost) escaping alpha particle, or also possibly by X-rays associated with the alpha emission process. It is interesting to note that the UKDMC experiment has reported that the anomalous events appear to present a summer/winter rate difference [27], attributed to some unknown systematic effect.

Our interpretation of the "bump events" in terms of surface nuclear recoils can be tested experimentally by using a surface implanted alpha source such as used by the Milano group [23]. In a forthcoming test run, we plan to measure simultaneous signals recorded by two adjacent germanium detectors to refine our analysis.

## Conclusions

The identification capabilities of our prototype detectors have allowed to identify four categories of events, instead of the two electron and nuclear recoil populations initially expected. One of these categories, which has been unnoticed by previous discrimination experiments, appears to be composed of surface nuclear recoils. This category of events may provide an explanation of the anomalous "bump" or "U" events observed by the UKDMC and Saclay WIMP experiments using NaI detectors. This interpretation can be tested experimentally in a relatively simple way.

It is then important to realize that the additional degrees of freedom represented by the surface nuclear and electron events, which present scintillation time responses differing from that of electron and nuclear volume interactions, must be included in any discrimination analysis of data recorded, e.g., in NaI experiments. A reanalysis of the experiments which have not taken into account these additional degrees of freedom in their analysis may then be necessary.

## Acknowledgments

We thank the technical staff of the LSM and of the participating laboratories for their invaluable help. This work has been partially funded by the EEC-Network program under contract ERBFMRXCT980167.

**References**

1. L. Baudis et al., Phys. Rev. D 59 (1999) 022001.
2. N. J. C. Spooner et al., Phys. Lett. B 273 (1991) 333.
3. R. Bernabei et al., Phys. Lett. B389 (1996) 757 ; Phys. Lett. B424 (1998) 195 ; Phys. Lett. B 450 (1999) 448.
4. P.F. Smith et al., Phys. Lett. B379 (1996) 299 ; Phys. Rep. 307 (1998) 275.
5. G. Gerbier et al., Astrop. Phys. 11 (1999) 287.
6. T. Shutt et al., Phys. Rev. Lett. 69 (1992) 3425.
7. R.W. Schnee et al., Phys. Rep. 307 (1998) 283.
8. D. Drain et al., Phys. Rep. 307 (1998) 297.
9. L. Berge et al., Nucl. Phys. B 70 (1999) 69-73.
10. D. L'Hôte et al., in Proc. 7$^{th}$ Int. Workshop et al. on Low Temperature Detectors, 27 July - 2 August 1997, Munich, Germany, pub. by MPI Physik, ISBN 3-00-002266-X.
11. M. Bravin et al., Astropart. Phys., 12 (1999) 107.
12. P. Meunier et al., Appl. Phys. Lett. 75 (1999) 1335.
13. A. de Bellefon et al., Astrop. Phys. 6 (1996) 35.
14. D. L'Hôte, X-F. Navick, R. Tourbot, J. Mangin and F. Pesty, J. Appl. Phys. 87 (2000) 1507.
15. P. Di Stefano et al., EDELWEISS collaboration, in preparation.
16. P.N. Luke, J. Appl. Phys. 64 (1988) 6858.
17. P. Di Stefano, Thèse de Doctorat, (Université Paris XI Orsay, 1998), unpublished.
18. G.E. Knoll, "Radiation Detection and Measurement", (J. Wiley, New York, 1989).
19. T. Shutt et al., in Proc. 7$^{th}$ Int. Workshop et al. on Low Temperature Detectors, 27 July - 2 August 1997, Munich, Germany, pub. by MPI Physik, ISBN 3-00-002266-X.
20. V. Chazal et al., Astrop. Phys. 9 (1998) 163.
21. F.M. Ipavic, R.A. Lundgren, B.A. Lambird and G. Gloeckler, Nucl. Instr. Methods 154 (1978) 291.
22. J.W. Zhou et al., Nucl. Instr. Methods Phys. Res. A 349 (1994) 225.
23. A. Alessandrello et al., Phys. Lett. B 408 (1997) 465.
24. G. Marie-Magdeleine et al., in Proceedings of the 8th International Workshop on Low Temperature Detectors, to appear in Nucl. Instr. Methods Phys. Res. A.
25. A. Broniatowski et al., in Proceedings of the 8th International Workshop on Low Temperature Detectors, to appear in Nucl. Instr. Methods Phys. Res. A.
26. V.A. Kudryavstev et al., Phys. Lett. B 452 (1999) 167.
27. N.J.C. Spooner et al., in "Proceedings of the 2$^{nd}$ International workshop on the Identification of Dark Matter", eds. N.J.C. Spooner and V. Kudryavstev, (World Scientific, Singapore, 1999).

**Figure captions:**

Fig. 1: Scatter diagram of the phonon amplitude vs. charge amplitude for a 70 g Ge detector exposed to a $^{60}$Co gamma source. The bias voltage for charge collection is 6 volts.

Fig. 2: Scatter diagram of the phonon amplitude vs. charge amplitude for a 70 g Ge detector exposed to a $^{57}$Co gamma source. The bias voltage for charge collection is 6 volts. Inset: incomplete charge collection events can be seen on the two 72.80 and 74.97 keV fluorescence Pb lines of the surrounding shield.

Fig. 3: Scatter diagram of the phonon amplitude vs. charge amplitude for a 70 g Ge detector exposed to a $^{252}$Cf neutron source. The bias voltage for charge collection is 6 volts.

Fig. 4: Scatter diagram of the recoil energy vs. charge amplitude for a prototype 70 g Ge detector resulting from a 1.17 kg x day background measurement. The scatter diagram exhibits four populations of events attributed to volume electron recoils, surface electron recoils, volume nuclear recoils, surface nuclear recoils (see text).

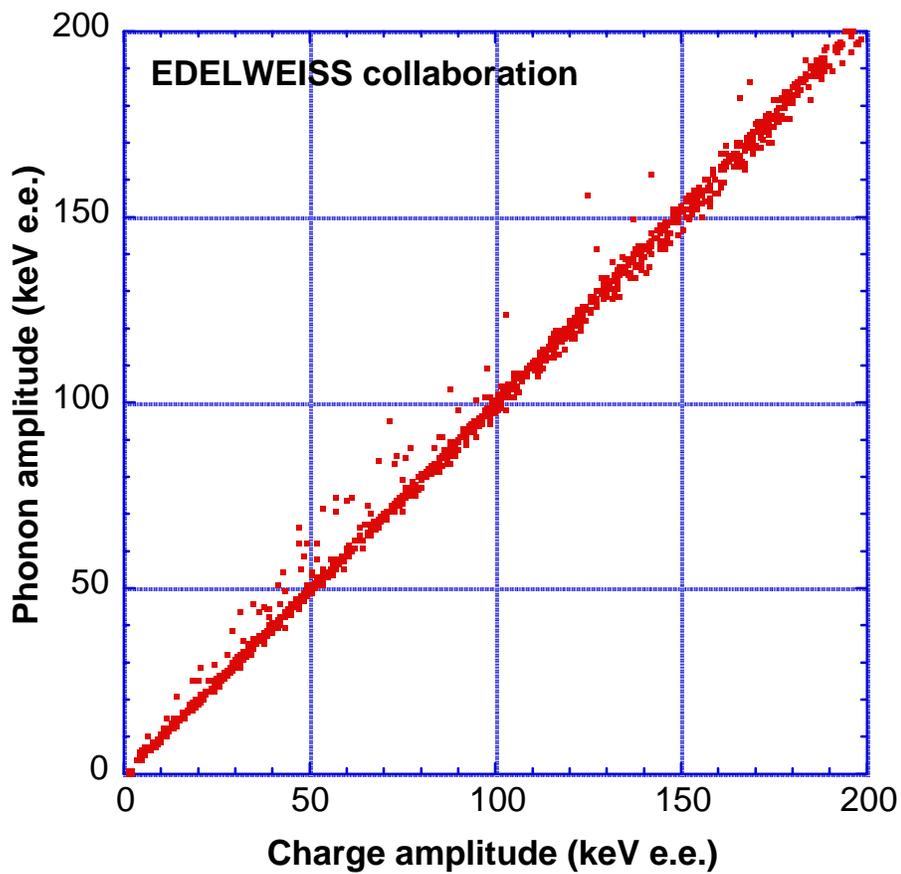

**Fig. 1**: Scatter diagram of the phonon amplitude vs. charge amplitude for a 70 g Ge detector exposed to a $^{60}$Co gamma source. The bias voltage for charge collection is 6 volts.

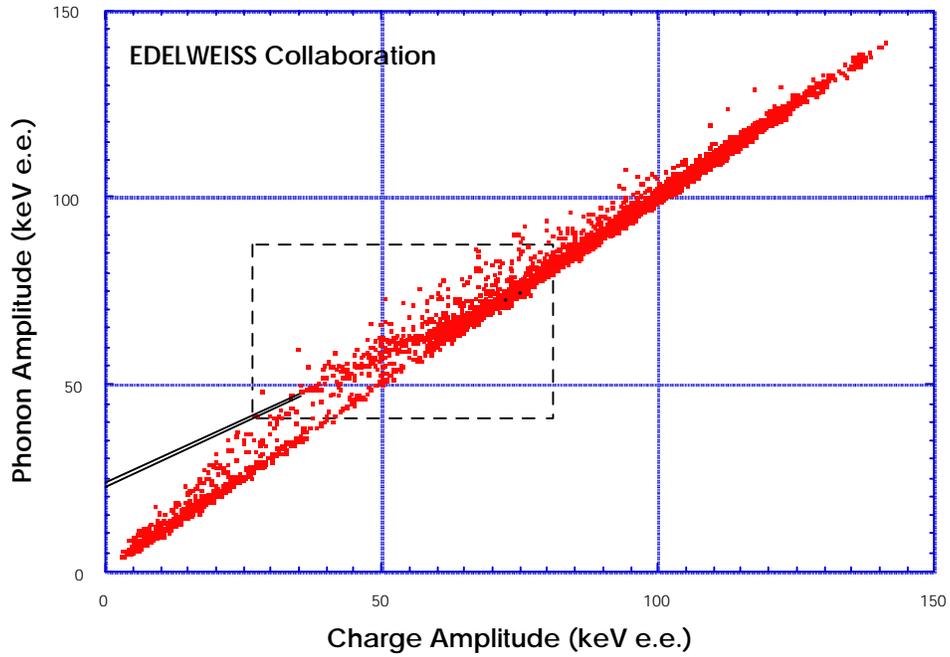

**Fig. 2**: Scatter diagram of the phonon amplitude vs. charge amplitude for a 70 g Ge detector exposed to a $^{57}$Co gamma source. The bias voltage for charge collection is 6 volts. Dashed box: incomplete charge collection events can be seen on the two 72.80 and 74.97 keV fluorescence Pb lines of the surrounding shield.

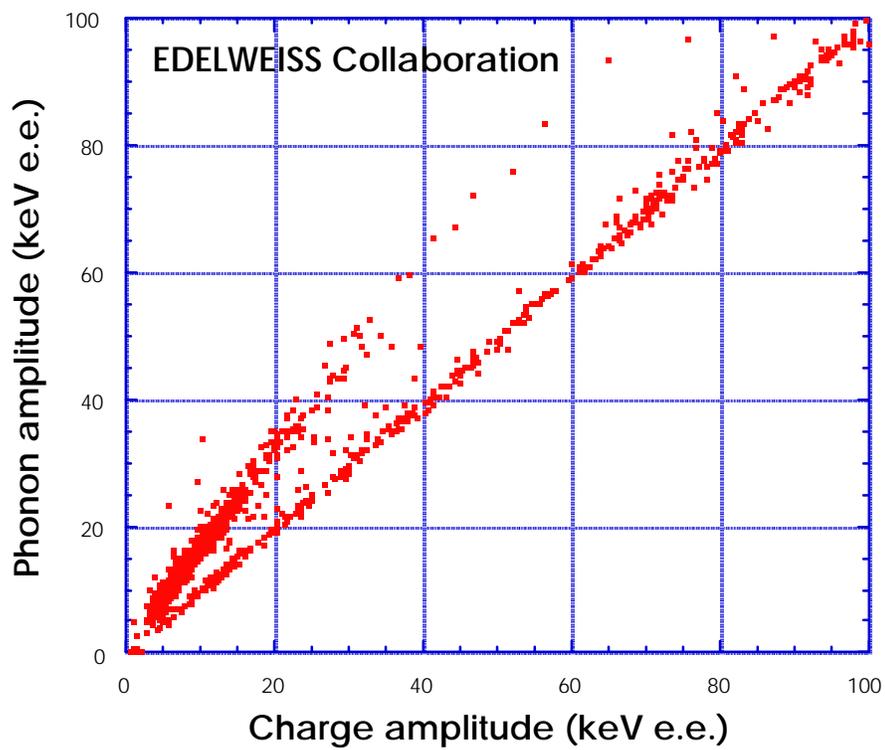

**Fig. 3**: Scatter diagram of the phonon amplitude vs. charge amplitude for a 70 g Ge detector exposed to a $^{252}$Cf neutron source. The bias voltage for charge collection is 6 volts.

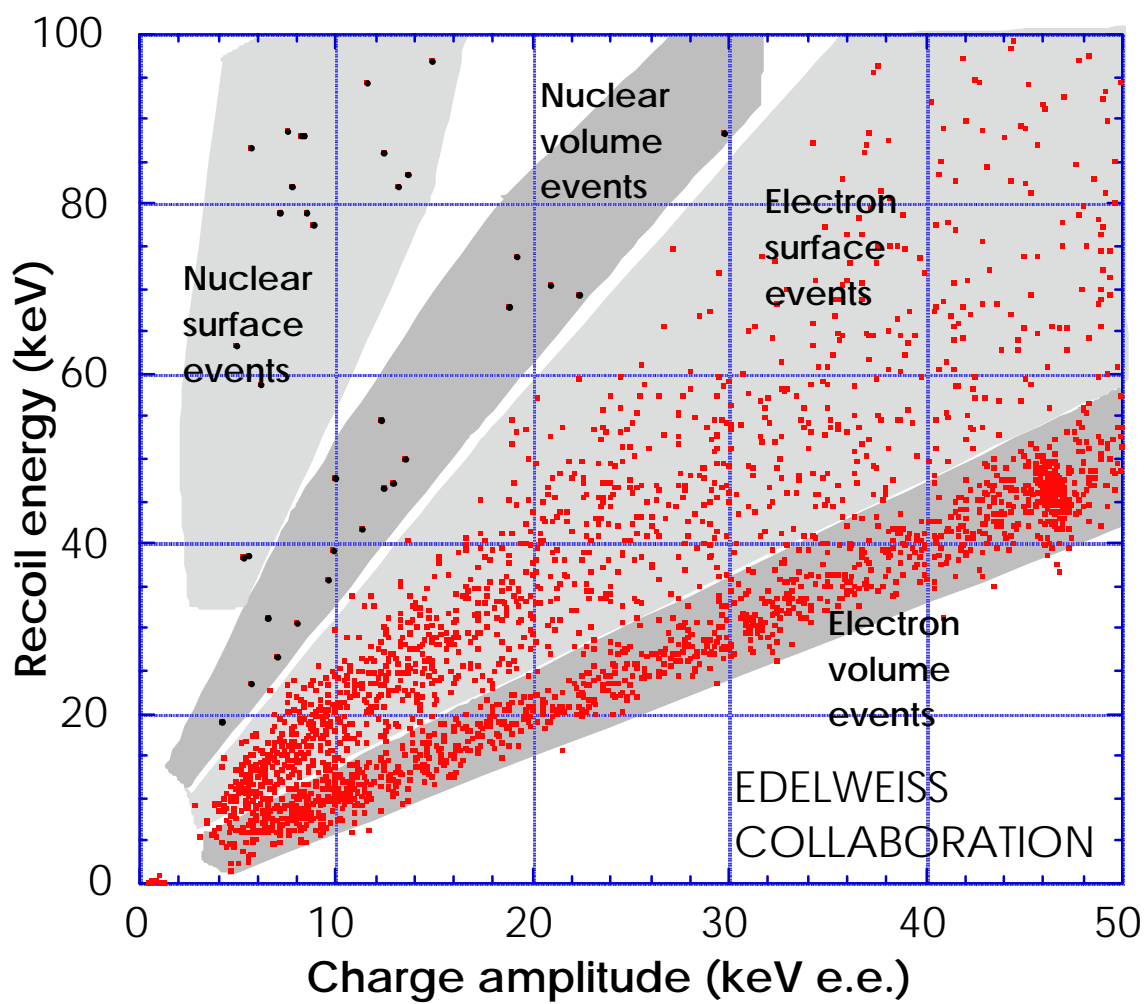

**Fig. 4**: Scatter diagram of the recoil energy vs. charge amplitude for a 70 g Ge detector resulting from a 1.17 kg x day background measurement. The scatter diagram exhibits four populations of events attributed to volume electron recoils, surface electron recoils, volume nuclear recoils, surface nuclear recoils (see text).